\documentclass[amssymb,aps]{revtex4}
\usepackage{graphicx}

\begin{document}

\title{Comment on ``Supersymmetry in the half-oscillator - revisited''}

\author{Ashok Das}
\affiliation{Department of Physics and Astronomy,
University of Rochester,
Rochester, NY 14627-0171, USA}
\author{S.~Pernice}
\affiliation{ Univ. del CEMA,
Buenos Aires, Argentina, 1054}
\bigskip

\begin{abstract}

We point out the flaw in the analysis of Gangopadhyaya and Mallow,
hep-th/0206133, where it is claimed that supersymmetry is broken in
the SUSY half-oscillator, even with a regularization respecting
supersymmetry. 

\end{abstract}

\maketitle

In an earlier paper \cite{das}, we had shown that supersymmetry, in
quantum mechanical theories with singular potentials or nontrivial
boundaries, is preserved if the system is regularized in a manner
respecting supersymmetry. This is in contrast to the earlier claims 
\cite{jevicki,roy,cooper} that supersymmetry is broken in such
systems. We  had shown, in particular, that if the superpotential is
regularized,  as opposed to the conventional wisdom of regularizing the
potential, supersymmetry is maintained in such systems in a natural
manner.  The reason for this is
quite clear. A regularized superpotential leads to a pair of
supersymmetric Hamiltonians for every value of the regularization
parameter 
whereas a conventionally regularized potential does not lead to a pair
of supersymmetric Hamiltonians for any value of the regularization
parameter. 

In a recent paper, Gangopadhyaya and Mallow \cite{asim} claim that of
the examples treated in \cite{das}, the supersymmetric half-oscillator
does not possess supersymmetry in spite of using the supersymmetric
regularizations proposed in \cite{das}. In fact, their analysis is
exactly the  same as given
in \cite{das}. However, their final solution of the boundary condition
and, therefore, their conclusion is
incorrect for a very simple reason that we wish to point out.

Let us recall that the SUSY half-oscillator was first studied in
\cite{roy} where the authors concluded that supersymmetry is
broken. Physically, their result can be understood as follows. If we
require the wavefunctions to vanish at the origin (as these authors do
and as would be the
case when one conventionally regularizes the potential), only odd
solutions of the harmonic oscillator are allowed. On the other hand,
supersymmetry requires the partner wavefunctions to have opposite
parity and since this is not allowed by the boundary conditions (which
follows from the particular regularization used),
supersymmetry must be broken. The problem with this conclusion, as
pointed  out in
\cite{das}, is that such a treatment of the system corresponds to
regularizing the potential in a manner which itself breaks
supersymmetry  and,
therefore, the result of such an analysis can be an artifact of the
regularization used. In stead, we had proposed to regularize the
superpotential directly as
\begin{equation}
W (x) = -\omega x \theta (x) + c \theta (-x)
\end{equation}
where $c$ is the regularization parameter and we take $c \rightarrow +
\infty $ at the end.  This superpotential leads to the
regularized supersymmetric potentials:
\begin{eqnarray}
V_+ (x)  & = & \frac{1}{2} [ (\omega^2 x^2 - \omega) \theta (x) + c^2 \theta
(-x) - c \delta (x) ]  \nonumber  \\
V_- (x)  & = & \frac{1}{2} [ (\omega^2 x^2 + \omega) \theta (x) + c^2 \theta
(-x) + c \delta (x) ]
\end{eqnarray}
The crucial difference between this and the earlier
treatment of this problem \cite{roy} is the appearance of Dirac delta
functions  in the potentials.
The delta function is attractive for one of the Hamiltonians while it
is repulsive for the other.
It is clear, therefore, that, in the presence of an attractive delta
function potential, the usual argument for the non existence of even
solutions is not automatic. In fact, we had shown explicitly that
supersymmetry is maintained in the limit $c\rightarrow +\infty$ in
this theory.  We had also proposed an alternate regularization in
\cite{das} 
that avoids the appearance of delta functions, and had shown that  the
end  result is the same if the regularization maintains supersymmetry.

It is surprising, therefore, that Gangopadhyaya and Mallow
\cite{asim}, who carry out exactly our analysis would
arrive at 
a different conclusion, namely, that supersymmetry is broken in this
system. Let us note that the boundary condition (eq. (23) in
\cite{das} and eq. (10) in \cite{asim}) is given by (we choose
$\omega = \frac{1}{2}$ for simplicity)
\begin{equation}
- \sqrt{2}\;\frac{\Gamma(\frac{1}{2}-\epsilon)}{\Gamma (-\epsilon)} =
\sqrt{c^{2}-2\epsilon} - c\label{boundary}
\end{equation}
where $\epsilon$ represents the energy of the
eigenstates. Gangopadhyaya and Mallow claim that $\epsilon = 0$ is not
a solution of the above equation, in which case 
supersymmetry will be broken. It is, of course, obvious that $\epsilon
= 0$ is a solution of (\ref{boundary}) for any value of $c$ and, in
particular, for $c\rightarrow +\infty$, but let us look at this more
carefully to understand the flaw in the analysis of \cite{asim}.

Let us define, for simplicity,
\begin{equation}
F(\epsilon) \equiv - \sqrt{2}\; \frac{ \Gamma (\frac{1}{2} - \epsilon)}{\Gamma
(-\epsilon)} \qquad {\mathrm{and}} \qquad  G(\epsilon, c) \equiv \sqrt{c^2 - 2
\epsilon} - c \quad ,
\end{equation}
so that we can represent the boundary condition (\ref{boundary}) as
\begin{equation}
F(\epsilon) = G (\epsilon,c)\label{boundary1}
\end{equation}
In a quantum theory, the boundary condition (\ref{boundary1}) may not
hold for all values of $\epsilon$.  The fact that energy is quantized means
that it can be satisfied only for some discrete values of
$\epsilon=\epsilon_{n}$.   Let us, in
particular, investigate whether $\epsilon = 0$ is a solution of the
equation, and if it is, whether it remains a solution
in the limit $c \rightarrow +\infty$.

Let us note that both $F(\epsilon)$ and $G(\epsilon,c)$, as complex
functions  of $\epsilon$, 
are analytic around $\epsilon=0$.  For $F(\epsilon)$, we can see this
from the fact
that the function $\frac{1}{\Gamma (-\epsilon)}$ is entire in the
whole $\epsilon$ 
plane with simple zeroes at the points $\epsilon = n$, $(n = 0,1,2,...)$,
and the function $\Gamma (\frac{1}{2} - \epsilon)$ is also analytic around
$\epsilon=0$,
with $\Gamma (\frac{1}{2}) = \sqrt{\pi}$. For $G(\epsilon,c)=
\sqrt{c^2 - 2 \epsilon} -c $, the same is also true.

Since both these functions are analytic around $\epsilon = 0$, a
Taylor  expansion around this value leads to,
\begin{eqnarray}
F(\epsilon) & \simeq & 0 + \sqrt{2 \pi} \, \epsilon +
{\mathcal{O}} (\epsilon^2) \label{F}\\
G(e,c)& \simeq & 0 - \frac{1}{c} \, \epsilon + {\mathcal{O}}
(\epsilon^2)\label{G}
\end{eqnarray}
The fact that $\epsilon = 0$ is a solution of equation
(\ref{boundary1})  is a consequence of 
the fact that the zeroth order terms in the Taylor expansion in
equations (\ref{F}) and (\ref{G}) coincide for that value.  Note that
this  result is independent of $c$ and, therefore, also holds in the limit $c
\rightarrow + \infty$.

The mistake in the analysis of Gangopadhyaya and Mallow lies in the
fact that,  for
whatever reason, they impose that the 
coefficient of the linear terms in the Taylor expansion should match, in
addition  to the zeroth order terms. Indeed, 
the equation below equation (11) (for $\omega = \frac{1}{2}$) in \cite{asim} is
precisely the requirement of the equality for the linear terms in
equations (\ref{F}) 
and (\ref{G}). This imposition, however, is unjustified, since that
would tantamount to requiring
\begin{equation}
F(0)=G(0,c),\qquad {\rm and}\qquad  F'(0)=G'(0,c)
\end{equation}
The boundary condition (\ref{boundary1}) does not have to hold for all
values of 
$\epsilon$, not even in a neighborhood of $\epsilon=0$. It should 
hold only  for discrete values of
$\epsilon$.  For the case $\epsilon=0$, this translates into the fact
that only the
zeroth order terms in the Taylor expansion of $F$ and $G$ should
be equal, but no extra conditions must be imposed on higher order
terms. The incorrect conclusion in \cite{asim} results from the
imposition of this additional condition. Supersymmetry is, in fact,
unbroken in this theory, when analyzed properly.
\vspace{1cm}

\noindent{\bf Acknowledgment:}

This work was supported in part by US DOE Grant number DE-FG 02-91ER40685.

\end{document}